\documentclass[onecolumn,preprintnumbers,prb,fleqn]{revtex4}
\usepackage{graphicx,color}
\usepackage{amsmath,amssymb}
\newcommand{\be}{\begin{equation}} 
\newcommand{\ee}{\end{equation}} 
\newcommand{\bea}{\begin{eqnarray}} 
\newcommand{\eea}{\end{eqnarray}}
\newcommand{\figI}
{\begin{figure}[htbp]
       \centering
       \includegraphics[width=6cm]{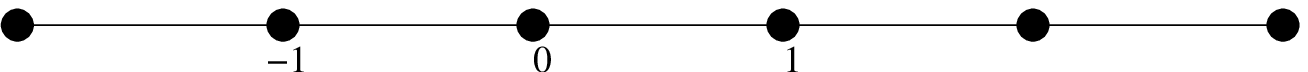}
	\caption{Monatomic Chain.}
	\end{figure}}
\newcommand{\figsqdisp}
{\begin{figure}[h]
       \includegraphics[width=6.5cm]{diat-chain.eps}
		\caption{Phonon spectrum of a diatomic chain.}
	\end{figure}}
\newcommand{\figSqM}
{\begin{figure}[h]
       \centering
       \includegraphics[width=4cm]{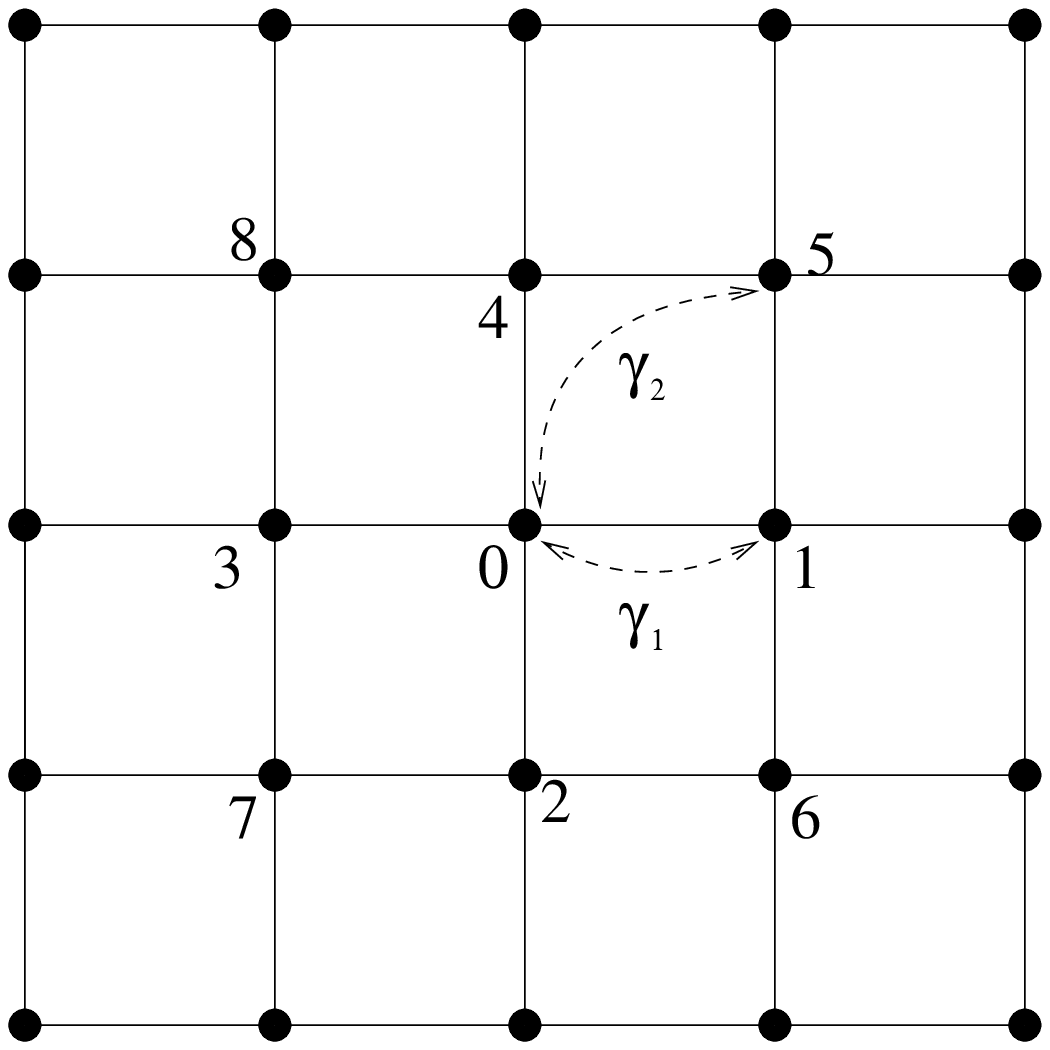}
	\caption{Monatomic square lattice.}
	\end{figure}}
\newcommand{\figsqBz}
{\begin{figure}[h]
       \centering
       \includegraphics[width=2.5cm]{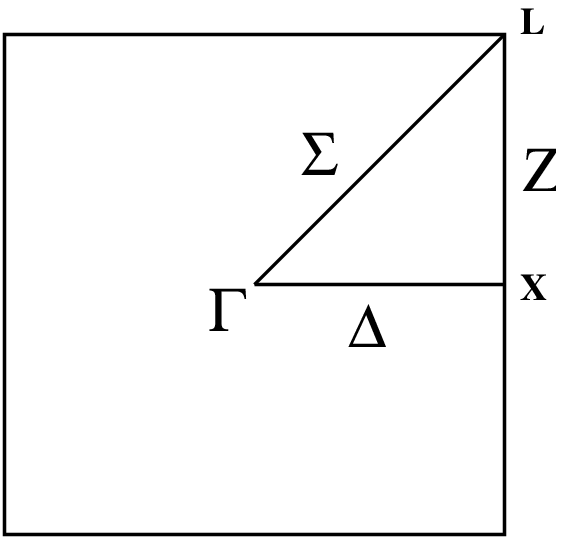}
	\caption{Brillouin Zone of square lattice.}
	\end{figure}}
\newcommand{\figmondisp}
{\begin{figure}[h]
      \centering
       \includegraphics[width=5cm]{sqrmonatomic.eps}
        \caption{Vibration spectrum of square lattice.}
	\end{figure}}
\newcommand{\figdiaeqm}
{\begin{figure}[h]
       \centering
       \includegraphics[width=6cm]{diato_equ_mass.eps}
	\caption{Vibration spectrum of diatomic square lattice with equal masses in the basis and with n. n. and n. n. n coupling.}
	\end{figure}}
\newcommand{\figdiauneqm}
{\begin{figure}[h]
       \centering
       \includegraphics[width=6cm]{diato_neq_mass.eps}
	\caption{Vibration spectrum of diatomic square lattice with unequal masses in the basis and with n. n. and n. n. n. coupling.}
	\end{figure}}
\newcommand{\figmonatvib}
{\begin{figure}[h]
       \centering
       \includegraphics[width=2.0 in]{monatomic.eps}
	\caption{Vibration spectrum of a one dimensional linear chain.}
	\end{figure}}
\newcommand{\figdiachain}
{\begin{figure}[h]
       \centering
       \includegraphics[width=2.0in]{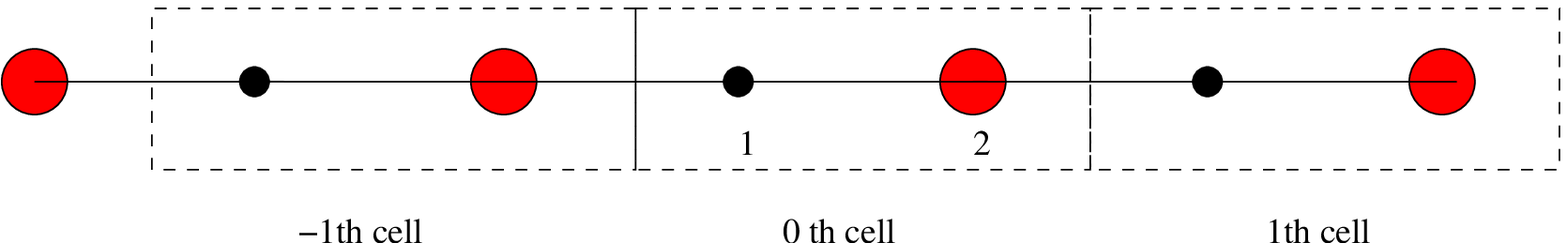}
	\caption{Diatomic Chain.}
	\end{figure}}
\newcommand{\figdiaeqmass}
{\begin{figure}[h]
       \centering
       \includegraphics[width=6cm]{diato_equ_mass1par.eps}
	\caption{Vibration spectrum of diatomic square lattice with equal masses in the basis.}
	\end{figure}} 
\newcommand{\figdiasqr}
{\begin{figure}
       \centering
       \includegraphics[width=4cm]{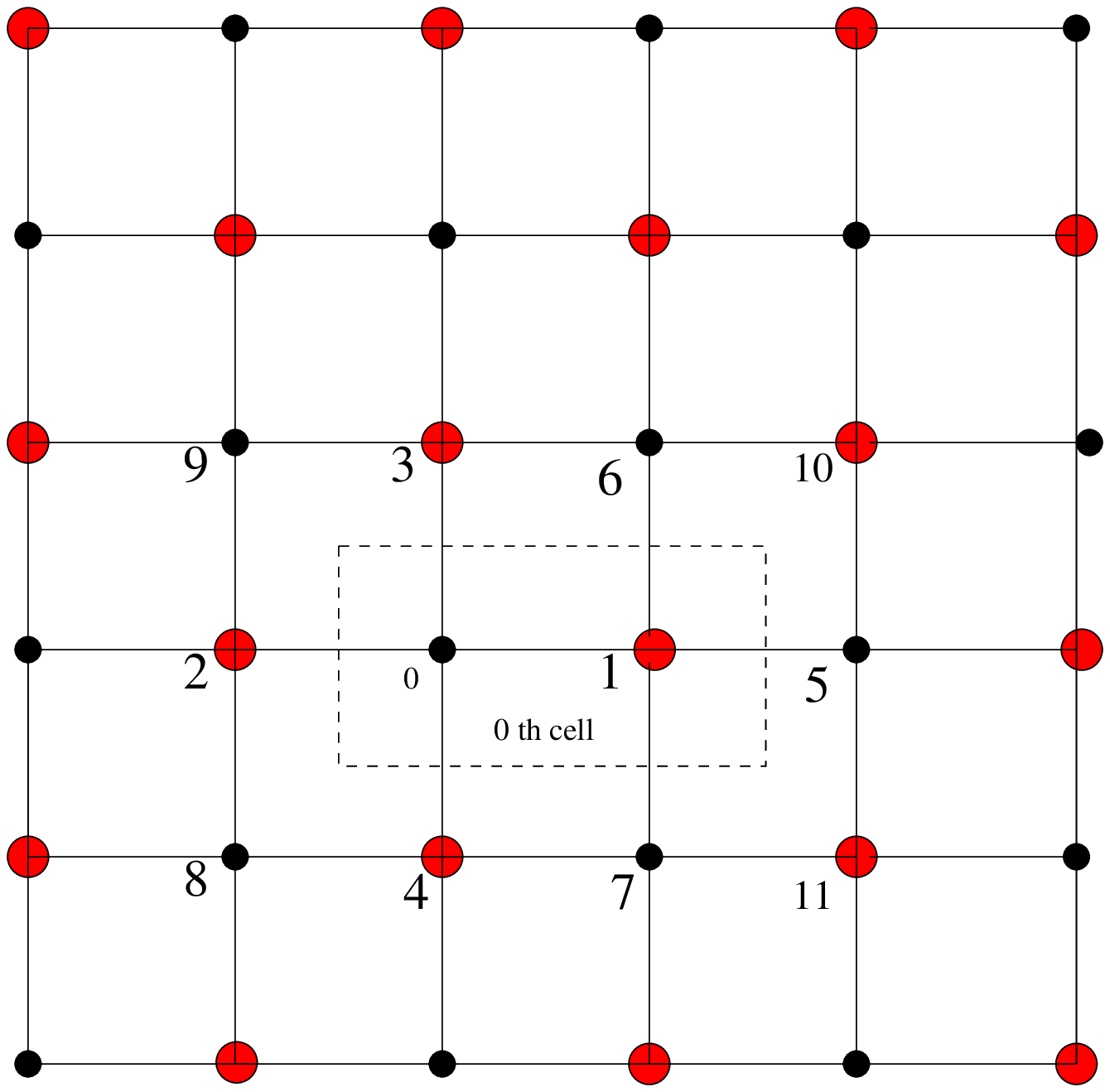}
	\caption{Diatomic square lattice.}
	\end{figure}}
\newcommand{\figmontrian}
{\begin{figure}[h]
       \centering
       \includegraphics[width=4.5cm]{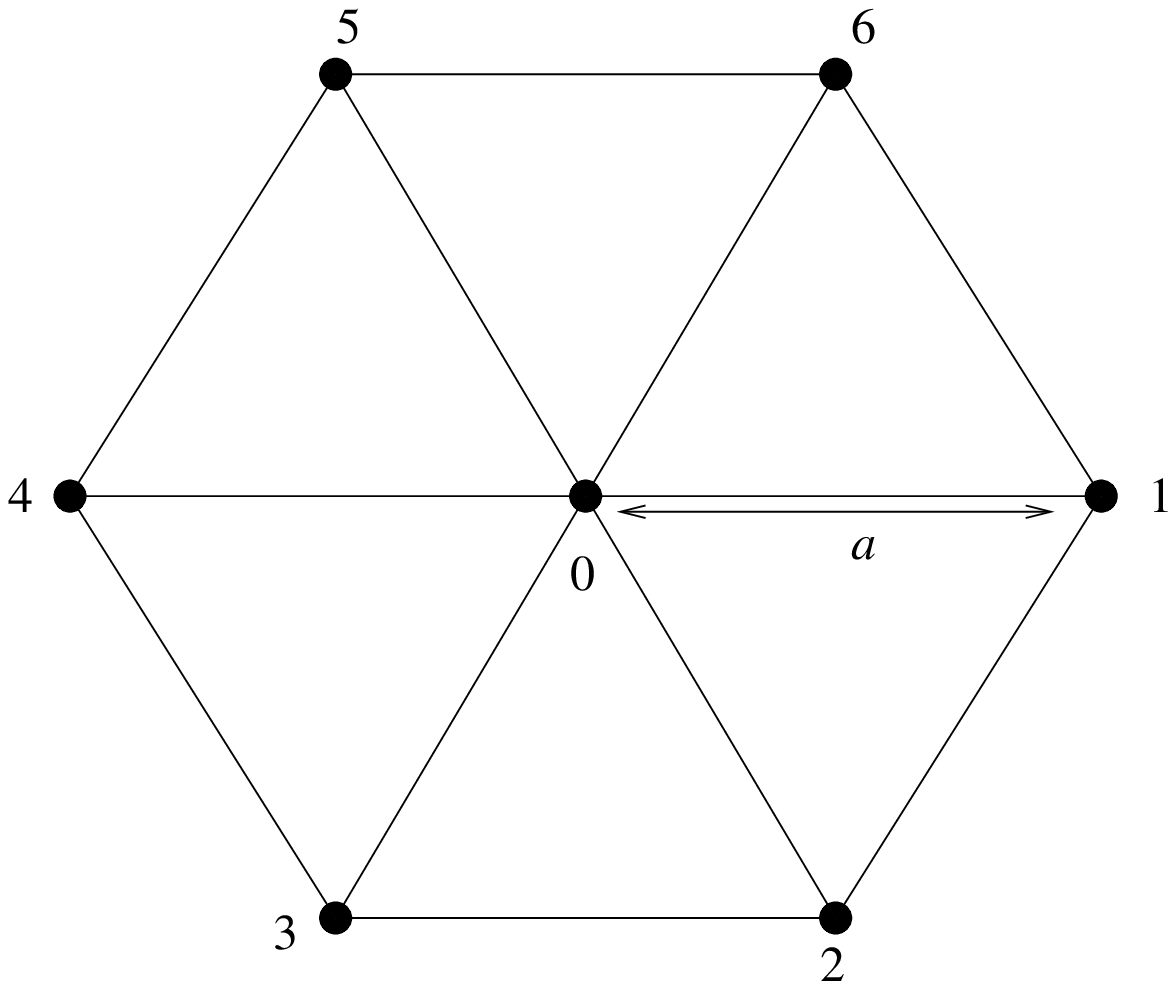}
        \caption{Triangular lattice.}
	\end{figure}} 
\newcommand{\fighexabz}
{\begin{figure}[h]
       \centering
       \includegraphics[width=2.5 cm]{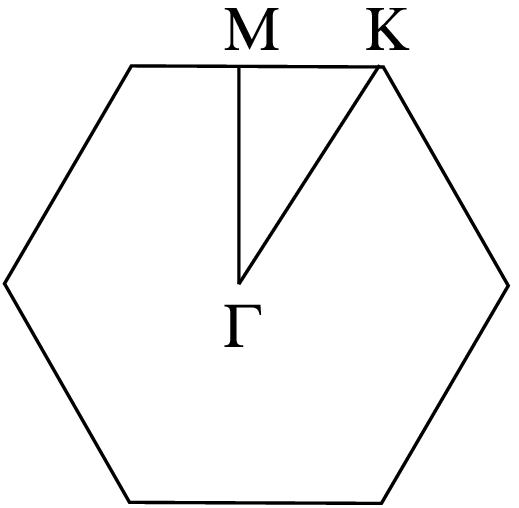}
        \caption{Brillouin Zone of Triangular lattice.}
	\end{figure}}
\newcommand{\figtridisp}
{\begin{figure}[h]
       \centering
       \includegraphics[width=5cm]{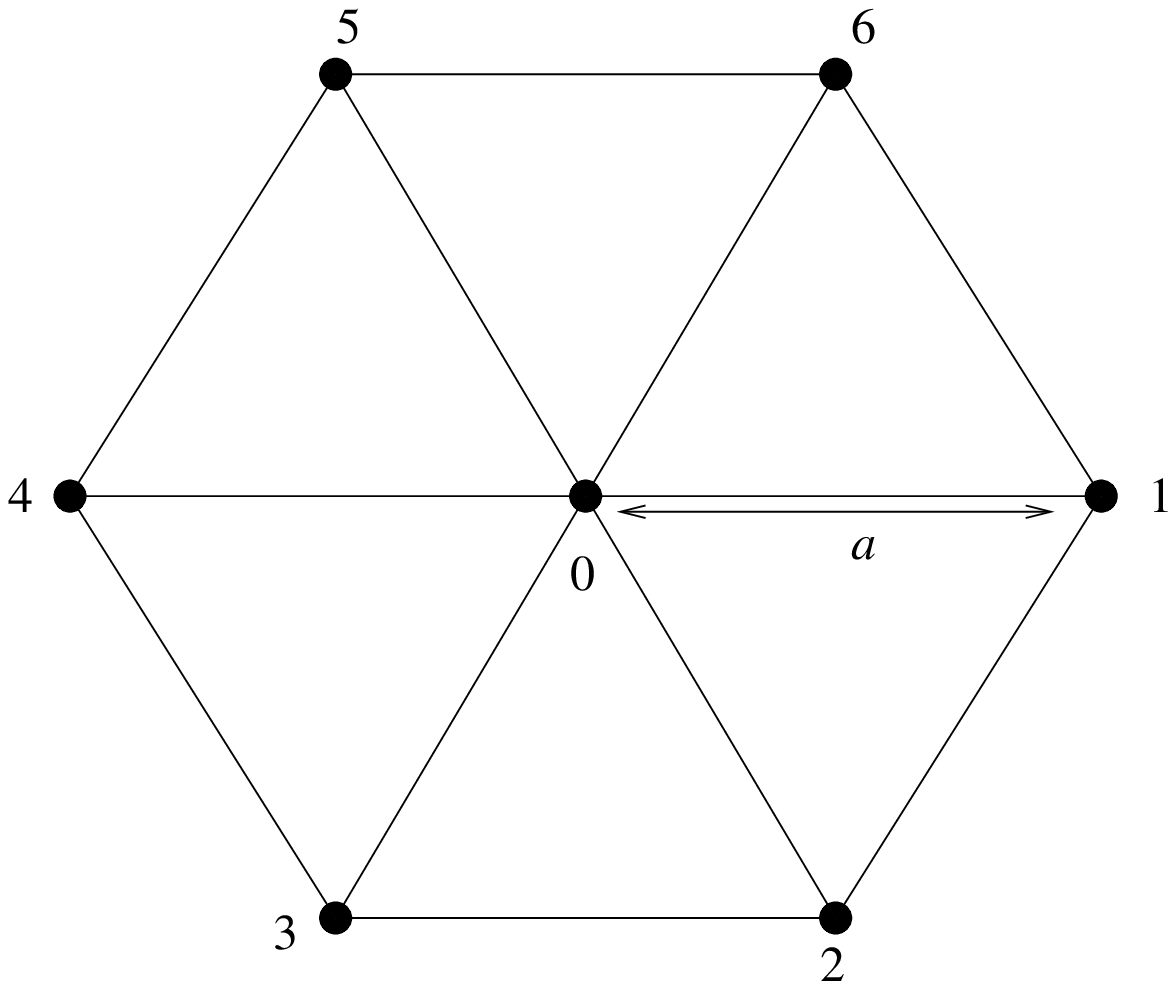}
        \caption{Vibration Spectrum of Triangular lattice.}
	\end{figure}}
\newcommand{\figgraphenestr}
{\begin{figure}[h]
       \centering
       \includegraphics[width=10cm]{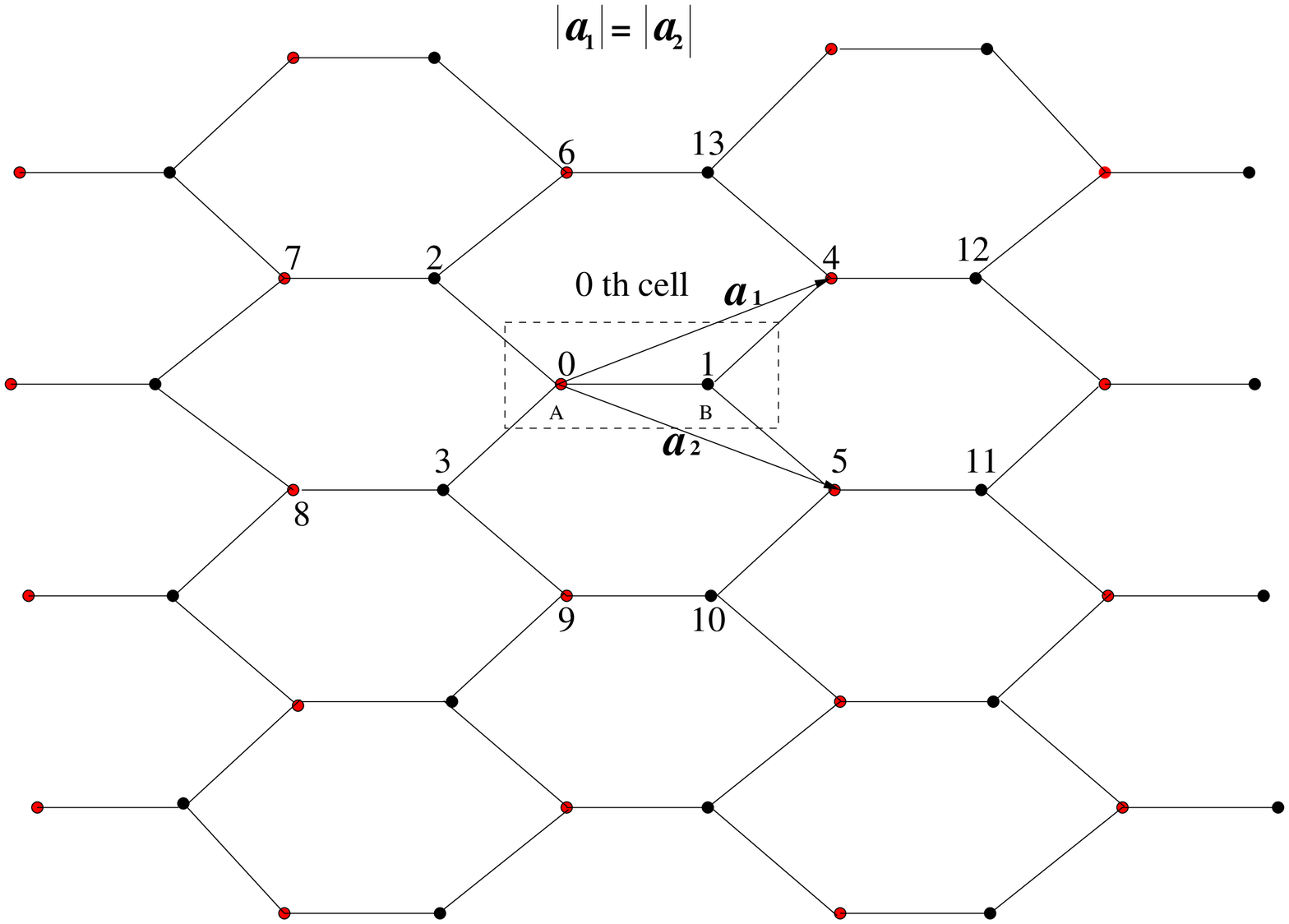}
	\caption{Structure of graphene sheet.}
	\end{figure}}
\newcommand{\figonepar}
{\begin{figure}[h]
      \centering
       \includegraphics[width=2.in]{phonon-final1par.eps}
	\caption{Phonon dispersion in graphene with n.n coupling only.}
	\end{figure}}
\vspace*{2cm}
\newcommand{\figgrphbz}
{\begin{figure}[h]
     \centering
         \includegraphics[width=2.5 cm]{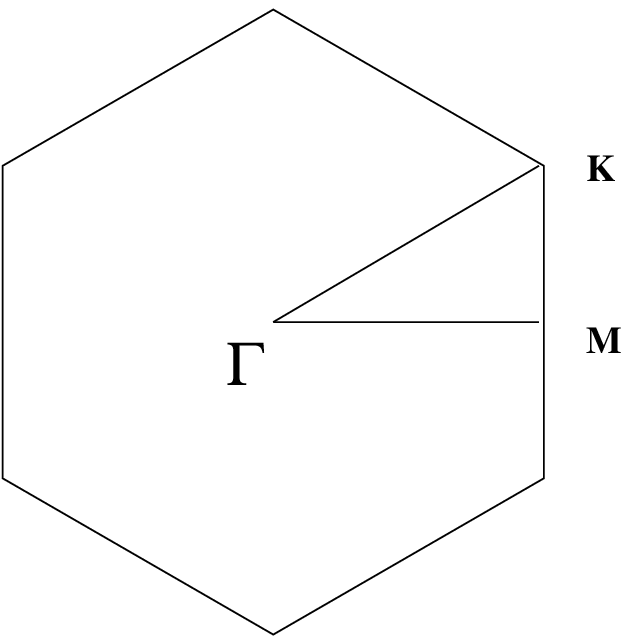}
	\caption{Brillouin zone of graphene.}
\end{figure}}
\newcommand{\figtwopar}
{\begin{figure}[h]
       \centering
        \includegraphics[width=2.in]{phonon_final2par.eps}
	\caption{Phonon dispersion in graphene along with n.n.n coupling.}
	\end{figure}}
\begin{document}
\title{Towards Phonon Spectrum of Graphene}
\author{Rupali Kundu}
\email{rupali@iopb.res.in}
\affiliation{ Institute of Physics, Bhubaneswar 751005 India.\\}
\date{\today}
\begin{abstract}
It is shown that the nearest neighbor coupling between the carbon atoms is not enough to reproduce the phonon spectrum as observed in graphene, the second neighbor force constant is essential. For completeness we have rederived the phonon spectrum of one dimensional chains and two dimensional lattices like square lattice and triangular lattice with nearest and next nearest neighbor coupling among the atoms. This article is essentially an account of the technical details of a review talk on phonon spectrum of graphene presented at Institute of Physics.
\end{abstract}
\maketitle
\section{Introduction}
Graphene is a one atom thick single layer of hexagonally arranged carbon atoms which has become experimentally accessible very recently. In this review only the in-plane vibration spectrum of perfect graphene has been calculated. The motivation was whether it is possible to find out the relevant features of phonon dispersion considering only nearest neighboring coupling among atoms because the important aspect of electronic dispersion in graphene is obtained with nearest neighbor hopping only, particularly the linear dispersion near the Brillouin zone corner ($K$ point). Also in the electron-phonon coupling in this system only those electrons around the $K$ point are going to take part. So it is relevant to see the nature of the vibration spectrum around the $K$ point. In section(II) the preliminaries of lattice vibration within harmonic approximation have been discussed, then in section(III) we discuss about vibration spectra of standard $1D$ monatomic 
linear chain, diatomic chain,  $2D$ square lattice without basis and with basis, triangular lattice and finally the graphene. Section(IV) compares the results obtained here for graphene vibration with the results calculated by L. M. Woods and G. D. Mahan \cite{MAHAN} and L. A. Falkovsky \cite{FALK}.
\section{some preliminaries}
The Hamiltonian of a vibrating crystal is given by
\be	
H = \sum_{nli}\frac{M_n}{2}  \dot u_i^2(n,l) +\frac{1}{2} \sum_{nli} \sum_{ml^{\prime}j} \phi_{ij}\dbinom{m,n}{l,~l^{'}}u_i(n,l)u_j(m,l^{\prime}), ~\text{where}~ \phi_{ij}\dbinom{m,n}{l,~l^{'}}=\left[{\frac{\partial^{2} U}{\partial u_i(n,l) \partial u_j(m,l^{\prime})}}\right]_{0}.
\ee
The notations used here are standard, that is, $M_n$ is the mass of the $n^{th}$ atom, $u_i(n,l)$ is a small displacement of the $n^{th}$ atom in the $l^{th}$ cell along $i^{th}$ direction, $i,j$ represent components of Cartesian coordinate axes and $U$ is the ion-ion interaction potential. Moreover, $\phi_{ij}(m,n;l,l^{'})$ is defined as the force acting on the $n^{th}$ atom in the $l^{th}$ cell along $i^{th}$ direction due to a unit displacement of the $m^{th}$ atom in the $l^{\prime th}$ cell along $j^{th}$ direction. The equation of motion of the $n^{th}$ atom in the $l^{th}$ cell is given by
\be
\label{eq:1}
M_n \ddot {u}_i(n,l)= -\sum_{ml^{\prime}j}\phi_{ij}\dbinom{m,n}{l,~l^{'}} u_j(m,l^{\prime}).
\ee
From the translational symmetry of a crystal the small displacement can be written as
\be
u_i(n,l)= M^{-\frac{1}{2}}_n u_{in} e^{-\imath \left(\omega t -\vec k \cdot \vec R_n(l)\right)},
\ee
where $u_{in}$ is the amplitude of vibration along $i^{th}$ direction of the $n^{th}$ atom, $\omega$ is angular frequency, $k$ is wave vector, $\vec R_n(l)$ is the lattice translation vector and the factor $M^{-1/2}_n$ has been chosen for convenience in further calculation.
It follows that
\be
\omega ^2 u_{ni}= \sum_{m,j} D_{ij}(mn,k) u_{mj},~~\text{where}~~
\label{eq:2}
D_{ij}(mn,k)= (M_m M_n)^{-\frac {1}{2}} \sum_l \phi_{ij} \dbinom{m,n}{l,~l^{'}}e^{-\imath \vec k \cdot \vec R_n(l)}
\ee
is the dynamical matrix which contains all the informations regarding the vibration of the lattice. Since this matrix is hermitian [$D_{ij}$ = $D^{*}_{ji}$], the eigenvalues are real and they give the vibration spectrum. Non-trivial solutions for these set of equations will be obtained if
\be
\label{eq:3}
|D_{ij}(mn,k)-\omega ^2 \delta_{ij} \delta_{mn}|= 0.
\ee
We shall now discuss the properties of atomic force constant $\phi_{ij}(m,n;l,l^{'})$.
\begin{enumerate}
\item Because  $\phi_{ij}(m,n;l,l^{'})$ is a second differential,
\be 
\phi_{ij}\dbinom{m,n}{l,~l^{'}}=\phi_{ji}\dbinom{n,m}{l^{'},~l}.
\ee
\item Translational symmetry means that $\phi_{ij}(m,n;l,l^{'})$ depends only on $\vec R_m(l)$ - $\vec R_n(l^{\prime})$. Hence,
\be
\phi_{ij}\dbinom{m,n}{l,~l^{'}}=\phi_{ij}\dbinom{m,n}{0,~l},~~\text{where $0$ refers to an origin of coordinates.}
\ee
\item The potential energy and the force on a given atom should be invariant under rigid body displacement of the whole crystal. This requires that
\be
\label{eq:4}
\sum_{n,l}\phi_{ij}\dbinom{m,n}{0,~l}= 0.
\ee
\item If the crystal has inversion symmetry then
\be
\phi_{ij}\dbinom{m,n}{0,l}=\phi_{ij}\dbinom{m,n}{0,-l}.
\ee
\end{enumerate}
These properties are useful in finding out relations between various force constants. To calculate the atomic force constants, let $\vec u_j(m)$ be the relative displacement of the $m^{th}$ atom in the $j^{th}$ direction with respect to the $n^{th}$ atom. Then the force acting on the $n^{th}$ atom in $i^{th}$ direction due to displacement of $m^{th}$ atom only is 
\be
f_i = \gamma_m {e_i(m)}  \sum_{j} \hat{e}_j(m).\vec u_j(m),~~\text{where}~~
\ee
$\hat e(m)$ is the unit vector along $\vec R_m(l)$ and $\gamma_m$ the spring constant between $n^{th}$ and $m^{th}$ atoms.
Then the total force acting on the $n^{th}$ atom in $i^{th}$ direction due to all other atoms is 
\be
F_i=\sum_{ml}\gamma_m  e_i(m)\sum_{j} \hat {e}_j(m).\vec u_j(m).
\ee
comparing the above equation with equation(\ref{eq:1}) and keeping in mind the definition of  $\phi_{ij}(m,n;0,l)$, we see that
\be
\label{eq:5}
 \phi_{ij}\dbinom{m,n}{0,~l}=-\gamma _m e_i(m) e _j (m).
\ee
\section{One and two dimensional lattices}
\subsection{Monatomic chain}
We shall now consider the vibration of a monatomic linear chain shown in figure(1). Let `$a$' be the lattice constant and $\gamma$ the nearest neighbor spring constant.
\figI
%\noindent 
Here the relevant unit vectors are 
\be
\hat{e_1}= \hat{i}, ~~~\hat{e_2}=-\hat{i}.\nonumber
\ee
For one dimensional case the index $(ij)$ corresponds to ($xx$) only and with one atom per cell the atomic index (m,n) in unimportant. So the indices in the force constant $\phi_{ij}(m,n;0,l)$ may be reduced to $\phi(0,l)$, where `$l$' represents nearest neighbor cell. According to the definition in equation(\ref{eq:5}), the force constants are calculated as
\be
\phi(0,1) =  -\gamma ~e_{1x}~e_{1x}
          = -\gamma ~~\text {and}~~
\phi(0,-1)=-\gamma ~e_{2x}~e_{2x}
           = -\gamma. \nonumber
\ee
Hence from the sum rule in equation (\ref{eq:4}), the self force constant is given by
\be
\phi(0,0)= - \left[\phi(0,1) + \phi(0,-1)\right]
          = 2 \gamma.\nonumber
\ee
In this case the dynamical matrix is a $(1\times1)$ matrix and the element is determined following the definition in equation (\ref{eq:2}), i.e.,
\bea
D(k) &=& \frac{1}{M}\left[\phi(0,~0)+\phi(0,~1)e^{\imath k a} + \phi(0,~-1)e^{-\imath k a}\right]\nonumber\\
    &=& \frac{2 \gamma}{M}\left[1-\cos(k a)\right].\nonumber
\eea
The eigen frequencies from equation (\ref{eq:3}) are given by
\be
 \omega= 2 \sqrt{\frac{\gamma}{M}}\sin(k a/2).
\ee
The brillouin zone of a one dimensional chain is a line with first brillouin zone boundaries at $\pm (\pi/a)$. The dispersion curve is shown in fig. (2).
\vspace*{.3cm}
\figmonatvib
In the long wavelength limit the vibration spectrum starts from zero and goes linear; it corresponds to the acoustic branch.
\subsection{Diatomic chain}
Let us consider a diatomic chain shown in fig. (3) with lattice constant $2a$, where `$a$' is interatomic distance.
\figdiachain
$M_1$ and $M_2$ are the masses in the unit cell and $\gamma$ is the coupling constant between them. In this case the unit vectors are same as that of monatomic chain.
For linear chain the ($ij$) index may be omitted from $\phi_{ij}$. From equation (\ref{eq:5}), the force constants are 
\be
\phi\dbinom{1,~2}{0,~0}= \phi\dbinom{1,~2}{0,-1}= -\gamma.\nonumber
\ee
The symbol $\phi(1,2;0,~0)$ indicates force constant between first and second atoms in the $0^{th}$ cell and $\phi(1,2;0,-1)$ indicates the same between first and second atoms in the $0^{th}$ and $-1^{th}$ cell. The self force constant~$\phi(1,1;0,~0)$ for the first atom is calculated from the sum rule in equation (\ref{eq:4}) as
\bea
\phi\dbinom{1,~1}{0,~0}+\phi\dbinom{1~~2}{0,~0}+\phi\dbinom{1,~2}{0,-1}&=& 0,\nonumber\\
 ~\text {i.e.}~~~~~~~~~~   \phi\dbinom{1,~1}{0,~0}&=& 2 \gamma. \nonumber 
\eea
Concentrating on the second atom in the $0^{th}$ unit cell, we get
\be
\phi\dbinom{2,~1}{0,~1}= \phi\dbinom{2,~1}{0,~0}= -\gamma.\nonumber
\ee
Hence from the sum rule
\bea
\phi\dbinom{2,~2}{0,~0}+\phi\dbinom{2,~1}{0,~1}+\phi\dbinom{2,~1}{0,~0}&=& 0,\nonumber\\ 
~\text {i.e.}~~~~~~~~~~ \phi\dbinom{2,~2}{0,~0}&=& 2 \gamma.\nonumber
\eea
The dynamical matrix elements are
\bea
D(11)&=& \frac{2 \gamma}{M_1},~~~D(12) = -\frac{\gamma}{\sqrt{M_1 M_2}}\left(1+e^{\imath k_x a}\right),\nonumber\\
~~~D(22)&=& \frac{2 \gamma}{M_2}~~~\text{and}~~~ D(21) = -\frac{\gamma}{\sqrt{M_1 M_2}}\left(1+e^{-\imath k_x a}\right).\nonumber
\eea
The eigen frequencies from equation (\ref{eq:3}) are given by
\be
\omega^2= \gamma \left(\frac{1}{M_1}+\frac{1}{M_2}\right)\pm\gamma \sqrt{\left(\frac{1}{M_1}+\frac{1}{M_2}\right)^2 - \frac{4 \sin^2(k_x a)}{M_1 M_2}}.
\ee
The dispersion curves are  plotted graphically in fig.(4).
\vspace{.5cm}
\figsqdisp
Due to difference in masses between two atoms in a unit cell, there is a gap between the bands at the zone boundary $\pm(\pi/2a)$ . If the two masses are equal, then the situation is basically one dimensional monatomic chain. So in that case there will be no gap at the boundary and the corresponding result will be in folded zone scheme, which is equivalent to the dispersion curve shown in figure (2) for monatomic chain.
\subsection{MONATOMIC SQUARE LATTICE}
Now let us consider a two dimensional monatomic square lattice. The nearest neighbors are numbered as $1$ to $4$ and next near neighbors as $5$ to $8$. The nearest neighbor and next nearest neighbor coupling constants are $\gamma_1$ and $\gamma_2$ respectively.
\figSqM
The next step is to calculate a set of unit vectors for the different atoms and then the force constants can be determined. The unit vectors are
\bea
\hat{e_1}&=& \hat{i},~~~\hat{e_2}=- \hat{j},~~~\hat{e_3}=- \hat{i},\nonumber\\
\hat{e_4}&=& \hat{j},~~~\hat{e_5}=\frac {1}{\sqrt 2} \hat{i} + \frac {1}{\sqrt 2} \hat{j},
~~~\hat{e_6}=\frac {1}{\sqrt 2} \hat{i} - \frac {1}{\sqrt 2} \hat{j},\nonumber\\
\hat{e_7}&=& -\frac {1}{\sqrt 2} \hat{i} - \frac{1}{\sqrt 2} \hat{j},~~\text{and}~~~\hat{e_8}=-\frac  {1}{\sqrt 2} \hat{i} + \frac {1}{\sqrt 2} \hat{j}. \nonumber
\eea
As we have numbered the neighboring atoms, it is convenient to write the force constants as $\phi_{ij}(m,n)$ instead of $\phi_{ij}(m,n;0,~l)$.
According to equation (\ref{eq:5}), the force constants are 
\bea
\phi_{xx}(0,~1)&=& \phi_{yy}(0,~2)=\phi_{xx}(0,~3)=\phi_{yy}(0,~4)=-\gamma_1,\nonumber\\
\phi_{yy}(0,~1)&=& \phi_{xy}(0,~1)=\phi_{yx}(0,~1)=\phi_{xx}(0,~2)=\phi_{xy}(0,~2)=
\phi_{yx}(0,~2)= \nonumber\\
\phi_{yy}(0,~3)&=& \phi_{xy}(0,~3)=\phi_{yx}(0,~3)=  \phi_{xx}(0,~4)=\phi_{xy}(0,~4)=\phi_{yx}(0,~4)=0,\nonumber\\
\phi_{xx}(0,~5)&=& \phi_{xx}(0,~6)=\phi_{yy}(0,~5)=\phi_{yy}(0,~6)=\phi_{xy}(0,~5)=\phi_{yx}(0,~5)=-\frac{\gamma_2}{2},\nonumber\\
\phi_{xy}(0,~6)&=& \phi_{yx}(0,~6)=\frac{\gamma_2}{2},\nonumber\\
\phi_{xx}(0,~7)&=& \phi_{xx}(0,~8)=\phi_{yy}(0,~7)=\phi_{yy}(0,~8)=\phi_{xy}(0,~7)=\phi_{yx}(0,~7)=-\frac{\gamma_2}{2},\nonumber \\
\phi_{xy}(0,~8)&=& \phi_{yx}(0,~8)=\frac{\gamma_2}{2}.\nonumber
\eea
The self force constant for the $0^{th}$ atom is
\bea
\phi_{xx}(0,~0)&=& -[\phi_{xx}(0,~1)+\phi_{xx}(0,~2)+\phi_{xx}(0,~3)+\phi_{xx}(0,~4)+\phi_{xx}(0,~5)+\nonumber\\
&& \phi_{xx}(0,~6)+\phi_{xx}(0,~7)+\phi_{xx}(0,~8)]\nonumber\\
&=& 2\left(\gamma_1 + \gamma_2\right).\nonumber
\eea
Similarly,
\bea
\phi_{yy}(0,~0)&=& 2\left(\gamma_1 + \gamma_2\right).\nonumber
\eea
According to equation (\ref{eq:2}) the dynamical matrix elements are 
\bea
D_{xx}(k)&=& \frac{1}{M}[\phi_{xx}(0,~1) e^{-\imath k_x a}+\phi_{xx}(0,~3) e^{\imath k_x a}+\phi_{xx}(0,~5) e^{-\imath (k_x a + k_y a)}+\phi_{xx}(0,~7) e^{-\imath (-k_x a - k_y a)} +\nonumber\\
&& \phi_{xx}(0,~8) e^{-\imath (-k_x a +k_y a)}+\phi_{xx}(0,~6) e^{-\imath (k_x a - k_y a)}+\phi_{xx}(0,~0)]\nonumber\\
&=& \frac{2}{M}\left[\gamma_1(1-\cos k_x a)+\gamma_2(1-\cos k_x a \cos k_y a)\right].\nonumber
\eea
Similarly,
\bea
D_{xy}(k)= D^*_{yx}(k)&=& \frac{\gamma_2}{M}[-\cos(k_x a + k_y a) + \cos(k_x a - k_y a)]\nonumber\\
&=& \frac{2 \gamma_2}{M} \sin k_x a \sin k_y a ~~~~\text{and}~~~~\nonumber\\
D_{yy}(k)&=& \frac{2}{M}[\gamma_1(1-\cos k_y a)+\gamma_2(1-\cos k_x a \cos k_y a)].\nonumber
\eea
The square brillouin zone of a square lattice with high symmetry points $\Gamma(0,0)$, $X(\pi/a,0)$ and $L(\pi/a,\pi/a)$ is shown in fig.(6).
\figsqBz
The eigen frequencies from equation (\ref{eq:3}) considering atomic mass as unity along the symmetry directions $\Delta\equiv (\Gamma - X)$ are 
\bea
\omega^2_1&=& 2(\gamma_1 + \gamma_2) (1-\cos k_x a),\\
\omega^2_2&=& 2\gamma_2 (1-\cos k_x a);
\eea
along $\Sigma \equiv (\Gamma - L)$
\bea
\omega^2_1&=& 2\left[\gamma_1(1-\cos k_x a) + 2 \gamma_2 \sin^2 k_x a\right],\\
\omega^2_2&=& 2\gamma_1(1-\cos k_x a);
\eea
and along $Z \equiv (X - L)$
\bea
\omega^2_1&=& 2\left[2\gamma_1 + \gamma_2 (1+\cos k_y a)\right],\\
\omega^2_2&=& 2\left[\gamma_1 (1-\cos k_y a)+ \gamma_2 (1+\cos k_y a)\right].
\eea 
The vibration spectra along various symmetry directions are shown in fig.(7).
\vspace*{.3 cm}
\figmondisp
From the curve it is evident that phonon frequencies are degenerate at $\Gamma$ and $L$ points whereas they are non-degenerate at $X$ point.From the above equations it is evident that one of the phonon branches has zero frequencies along the $\Gamma - X$ direction in absence of second neighbor coupling. But under same condition  it has non-zero frequencies along $X-L$ and $\Gamma - L$ directions.
\subsection{SQUARE LATTICE WITH A BASIS}
Let us now consider a square lattice with a basis of two atoms of masses $M_1$ and $M_2$. The interatomic distance is `$a$', i.e., lattice periodicity is $2a$. In fig. (8) the two atoms in the $0^{th}$ cell have been numbered as $0$ and $1$. The atoms numbered as $1, 2, 3 ,4$ and $6, 7, 8, 9$ are the  nearest and next nearest neighbors respectively of the $0^{th}$ atom. The nearest and next nearest neighbors of the first atom are $0, 5, 6, 7$ and $3, 4, 11, 10$ respectively. Let $\gamma$, $\gamma_1$ and $\gamma_2$ be the coupling constants among nearest $M_1$, $M_2$; $M_1$, $M_1$ and $M_2$, $M_2$ respectively.
\figdiasqr
Proceeding in the same way like the monatomic square lattice, the different force constants become
\bea
\phi_{xx}(0,~1)&=& \phi_{xx}(0,~2)=\phi_{yy}(0,~3)=\phi_{yy}(0,~4)=-\gamma,\nonumber\\
\phi_{xx}(0,~3)&=& \phi_{xx}(0,~4)=\phi_{yy}(0,~1)=\phi_{yy}(0,~4)=0,\nonumber\\
\phi_{xy}(0,~1)&=& \phi_{xy}(0,~2)=\phi_{xy}(0,~3)=\phi_{xy}(0,~4)=0, \nonumber\\
\phi_{xx}(1,~0)&=& \phi_{xx}(1,~5)=\phi_{yy}(1,~6)=\phi_{yy}(1,~7)=-\gamma, \nonumber\\
\phi_{xx}(1,~6)&=& \phi_{xx}(1,~7)=\phi_{yy}(1,~0)=\phi_{yy}(1,~5)=0. \nonumber
\eea
The corresponding $\phi^{'s}_{xy}$ and $\phi^{'s}_{yx}$ are zero.
\bea
\phi_{xx}(0,~6)&=& \phi_{xx}(0,~7)=\phi_{xx}(0,~8)=\phi_{xx}(0,~9)=-\frac{\gamma_1}{2},\nonumber\\
\phi_{yy}(0,~6)&=& \phi_{yy}(0,~7)=\phi_{yy}(0,~8)=\phi_{yy}(0,~9)=-\frac{\gamma_1}{2}, \nonumber\\
\phi_{xy}(0,~6)&=& \phi_{xy}(0,~8)=\phi_{yx}(0,~6)=\phi_{yx}(0,~8)=-\frac{\gamma_1}{2}, \nonumber\\
\phi_{xy}(0,~7)&=& \phi_{xy}(0,~9)=\phi_{yx}(0,~7)=\phi_{yx}(0,~9)=\frac{\gamma_1}{2}.\nonumber
\eea
Hence from the sum rule in equation (\ref{eq:4}) the self force constants for the $0^{th}$ atom in the $0^{th}$ unit cell are
\bea 
\phi_{xx}(0,~0)&=& \phi_{yy}(0,~0)=2\gamma + 2 \gamma_1\nonumber~~~\text{and}~~~\\ 
\phi_{xy}(0,~0)&=& \phi_{yx}(0,~0)=0. \nonumber
\eea
Concentrating on the second atom in the $0^{th}$ cell, we get the following force constants
\bea
\phi_{xx}(1,~10)&=& \phi_{xx}(1,~4)=\phi_{xx}(1,~3)=\phi_{xx}(1,~11)=-\frac{\gamma_2}{2},\nonumber\\
\phi_{yy}(1,~10)&=& \phi_{yy}(1,~4)=\phi_{yy}(1,~3)=\phi{yy}(1,~11)=-\frac{\gamma_2}{2},\nonumber\\
\phi_{xy}(1,~10)&=& \phi_{xy}(1,~3)=\phi_{yx}(1,~10)=\phi_{yx}(1,~3)=-\frac{\gamma_2}{2},\nonumber\\
\phi_{xy}(1,~11)&=& \phi_{xy}(1,~4)=\phi_{yx}(1,~11)=\phi_{yx}(1,~4)=\frac{\gamma_2}{2}\nonumber.
\eea
The corresponding self force constants for the second atom are
\bea
\phi_{xx}(1,~1)&=& \phi_{yy}(1,~1)=2\gamma + 2 \gamma_2\nonumber~~~\text{and}~~~\\ 
\phi_{xy}(1,~1)&=& \phi_{yx}(1,~1)=0\nonumber.
\eea
The dynamical matrix elements are
\bea
D_{xx}(00)&=& \frac{1}{M_1}[\phi_{xx}(0,~6) e^{-\imath (k_x a + k_y a)}+\phi_{xx}(0,~8) e^{\imath (k_x a + k_y a)}+\nonumber\\
&&\phi_{xx}(0,~9) e^{\imath (k_x a - k_y a)}+\phi_{xx}(0,~7) e^{-\imath (k_x a - k_y a)}+\phi_{xx}(0,~0)]\nonumber\\
          &=& \frac{2}{M_1}\left[\gamma + \gamma_1(1-\cos k_x a \cos k_y a)\right].\nonumber
\eea
Similarly,
\bea 
D_{xy}(00)&=& -\frac{2 \gamma_1}{M_1}\sin k_x a \sin k_y a,~~
D_{yx}(00)= -\frac{2 \gamma_1}{M_1} \sin k_x a \sin k_y a,\nonumber\\
D_{yy}(00)&=& \frac{2}{M_1}[\gamma + \gamma_1(1-\cos k_x a \cos k_y a)],~~
D_{xx}(11)= \frac{2}{M_2}[\gamma + \gamma_2(1-\cos k_x a \cos k_y a)],\nonumber\\
D_{xy}(11)&=& -\frac{2 \gamma_2}{M_2}\sin k_x a \sin k_y a,~~
D_{yx}(11)= -\frac{2 \gamma_2}{M_2} \sin k_x a \sin k_y a,\nonumber\\
D_{yy}(11)&=& \frac{2}{M_2}[\gamma + \gamma_2(1-\cos k_x a \cos k_y a)],~~
D_{xx}(01)= -\frac{2 \gamma}{\sqrt{M_1 M_2}}\cos k_x a,\nonumber\\
D_{xy}(01)&=&D_{yx}(01) = 0,~~D_{yy}(01)=-\frac{2 \gamma}{\sqrt{M_1 M_2}}\cos k_y a,\nonumber \\
D_{xx}(10)&=& -\frac{2 \gamma}{\sqrt{M_1 M_2}}\cos k_x a,~~D_{xy}(10)=D_{yx}(10)= 0~~\text{and} ~~
D_{yy}(10)= -\frac{2 \gamma}{\sqrt{M_1 M_2}}\cos k_y a.\nonumber
\eea
The solutions to the secular equation (\ref{eq:3})
are plotted along the symmetry directions of the square brillouin zone with high symmetry points $\Gamma(0,0)$, $X(0,\pi/2a)$ and $L(\pi/2a,\pi/2a)$ for three different cases.\\
\underline {Case $1$}:
Two masses in the basis are same and $\gamma\neq {0.0}$ but $\gamma_1$=$\gamma_2=0$.
\vspace*{.2cm}
\figdiaeqmass
In this case both acoustic and optical modes are degenerate at $\Gamma$ point; at $X$ point two are degenerate and two are nondegenerate but at $L$ point all four are degenerate. Four branches are non-degenerate along $\Gamma-X$ direction. In this direction one of the acoustic branches is having vanishing frequencies and one optical branch is dispersionless. Along $X-L$ direction one acoustic and one optical branches are degenerate and along $\Gamma - L$ direction both the branches are degenerate.\\ 
\underline {Case $2$}:
Two masses in the basis are same, $\gamma\neq{0.0}$ and $\gamma_1=\gamma_2\neq{0.0}$.
\vspace*{.3cm}
\figdiaeqm
Here also both acoustic and optical modes are degenerate at $\Gamma$ point; at $X$ and $L$ points two are degenerate and two are nondegenerate. Four branches are non-degenerate along $\Gamma - X$ direction as well as along $X - L$ and $L - \Gamma$ directions.\\
\underline {Case $3$}:
Two masses in the basis are unequal and also $\gamma_1\neq\gamma_2\neq{0.0}$.
\vspace*{.2cm}
\figdiauneqm
Here both acoustic and optical modes are degenerate at $\Gamma$ point but at $X$ and $L$ points all are nondegenerate. All the four branches are nondegenerate along $\Gamma - X$, $X - L$ and $L - \Gamma$ directions.
\subsection{MONATOMIC TRIANGULAR LATTICE}
The structure of a triangular lattice is shown in the fig.(12). Each unit cell of this lattice consists of one atom.
\figmontrian
The nearest neighbors of the $0^{th}$ atom are numbered as $1$ to $6$ and $\gamma$ is the spring constant among the neighbors.
\bea
\hat{e_1}&=& \hat{i},~~~\hat{e_2}=\frac {1}{2} \hat{i} - \frac {\sqrt 3}{2} \hat{j},~~~\hat{e_3}= -\frac {1}{2} \hat{i} - \frac {\sqrt 3}{2} \hat{j},\nonumber\\
\hat{e_4}&=& -\hat{i},~~~\hat{e_5}=-\frac {1}{2} \hat{i} + \frac {\sqrt 3}{2} \hat{j}~~~\text{and} ~~~\hat{e_6}=\frac {1}{2} \hat{i} + \frac {\sqrt 3}{2} \hat{j} \nonumber
\eea
are the unit vectors with respect to the $0^{th}$ atom.
The force constants calculated by using equation (\ref{eq:5}) are
\bea
\phi_{xx}(0,~1)&=& \phi_{xx}(0,~4)=-\gamma,\nonumber\\
\phi_{xx}(0,~2)&=& \phi_{xx}(0,~3)=\phi_{xx}(0,~5)=\phi_{xx}(0,~6)=-\frac{\gamma}{4},\nonumber\\
\phi_{xy}(0,~1)&=& \phi_{yx}(0,~1)=\phi_{xy}(0,~4)=\phi_{yx}(0,~4)=\phi_{yy}(0,~1)=\phi_{yy}(0,~4)=0,\nonumber\\
\phi_{xy}(0,~3)&=& \phi_{xy}(0,~6)=\phi_{yx}(0,~3)=\phi_{yx}(0,~6)=-\frac{\sqrt{3} \gamma}{4},\nonumber\\
\phi_{xy}(0,~2)&=& \phi_{xy}(0,~5)=\phi_{yx}(0,~2)=\phi_{yx}(0,~5)=\frac{\sqrt{3} \gamma}{4},\nonumber\\
\phi_{yy}(0,~2)&=& \phi_{yy}(0,~3)=\phi_{yy}(0,~5)=\phi_{yy}(0,~6)=-\frac{3 \gamma}{4}.\nonumber
\eea
Hence from the sum rule in equation (\ref{eq:4}) the self force constants are $\phi_{xx}(0,~0)= 3 \gamma$, $\phi_{xy}(0,~0)=\phi_{yx}(0,~0)=0$ and $\phi_{yy}(0,~0)=3 \gamma$. The elements of the dynamical matrix are
\bea
D_{xx}(k)&=& \frac{\gamma}{M}\left[3-2 \cos k_x a - \cos \frac{k_x a}{2} \cos \frac{\sqrt{3}k_y a}{2}\right],\nonumber\\
D_{xy}(k)&=& D^*_{yx}(k)=-\frac{\sqrt{3}\gamma}{M} \sin \frac{k_x a}{2} \sin \frac{\sqrt{3}k_y a}{2}~~~~\text{and}~~~~\nonumber\\
D_{yy}(k)&=& \frac{3 \gamma}{M}\left[1- \cos \frac{k_x a}{2} \cos \frac{\sqrt{3}k_y a}{2}\right].\nonumber
\eea
Figure(13) shows the hexagonal brillouin zone of triangular lattice with high symmetry points $\Gamma(0,0)$, $M(0,2\pi/\sqrt{3}a)$ and $K(2\pi/3a, 2\pi/\sqrt{3}a)$.
\fighexabz
Considering atomic mass to be unity, the dispersion relations along the symmetry directions {$\Gamma - M$} are
\bea
\omega^2_1&=& \gamma\left[1-\cos \frac{\sqrt{3}k_y a}{2}\right], \\
\omega^2_2&=& 3 \gamma\left[1-\cos \frac{\sqrt{3}k_y a}{2}\right];
\eea
along {$M - K$} are
\bea
\omega^2_1&=& 3 \gamma\left[1+\cos \frac{k_x a}{2}\right], \\
\omega^2_2&=& \gamma\left[3-2 \cos k_x a+\cos \frac{k_x a}{2}\right];
\eea
and along {$K - \Gamma$} are
\be
\omega^2_1= \omega^2_2= \gamma \left[ (3-\cos k_x a - 2 \cos \frac{k_x a}{2} \cos \frac{\sqrt{3}k_y a}{2}) \pm \sqrt{(\cos \frac{k_x a}{2} \cos\frac{\sqrt{3}k_y a}{2})^2 + 3 \sin^2\frac{k_x a}{2}\sin^2\frac{\sqrt{3}k_y a}{2}}~~\right].
\ee
The graphical plot of the vibration spectrum is in fig.(14).
\vspace{.3cm}
\figtridisp
Both the vibration spectra start from zero at the $\Gamma$ point (acoustic branch). They are nondegenerate at $M$ point but degenerate at $K$ point. The nearest neighbor coupling among atoms is sufficient to produce the vibration spectrum over the whole brillouin zone.\\
\subsection{LATTICE VIBRATION OF HEXAGONAL GRAPHENE SHEET}
The structure of hexagonal graphene sheet is shown in fig.(15).
\figgraphenestr
It consists of two sublattices (say $A$ and $B$) where $A$ and $B$ type of carbon atoms differ by their bond orientations. Thus each primitive cell contains two atoms. In the figure the two atoms in the $0^{th}$ cell have been numbered as $0$ and $1$. The nearest neighbors (n.n.) of one kind of atom (e.g. $A$) are three atoms belonging to other sublattice ($B$) and the next nearest neighbors (n.n.n.) are six atoms in the same sublattice. The atoms numbered as $1, 2, 3 $ and $4, 5, 6, 7, 8, 9$ are the  nearest and next nearest neighbors respectively of the $0^{th}$ atom. The nearest and next nearest neighbors of the first atom are $0, 4, 5$ and $2, 3, 10, 11, 12, 13$ respectively. Let us now connect the n.n. and n.n.n. atoms by springs of spring constant $\gamma$ and $\gamma_1$ respectively. The distance between adjacent carbon atoms in the plane is $a_0 (1.42\AA)$. The magnitude of the primitive vectors $(\vec a_1$ and $\vec a_2)$ is `$a$'$(=a_0 \sqrt 3)$. The set of unit vectors towards n.n. and n.n.n. with respect to the atom $A$ (at $0$) are the following
\bea
\hat{e_1}&=& \hat{i},~~~\hat{e_2}=-\frac {1}{2} \hat{i} + \frac {\sqrt 3}{2} \hat{j},~~~\hat{e_3}=-\frac {1}{2} \hat{i} - \frac {\sqrt 3}{2} \hat{j},\nonumber\\
\hat{e_4}&=& \frac {\sqrt 3}{2} \hat{i} + \frac {1}{2} \hat{j},~~~\hat{e_5}=\frac {\sqrt 3}{2} \hat{i} - \frac {1}{2} \hat{j},~~~\hat{e_6}=\hat{j},\nonumber\\
\hat{e_7}&=& -\frac {\sqrt 3}{2} \hat{i} + \frac {1}{2} \hat{j},~~~\hat{e_8}=-\frac {\sqrt 3}{2} \hat{i}-\frac {1}{2} \hat{j}~~\text{and}~~\hat{e_9}=-\hat{j}.\nonumber
\eea
Similarly, the unit vectors with respect to the first atom can be calculated.
According to the definition in equation (\ref{eq:5}), the force constants among nearest $A-B$ and $A-A$ type atoms are calculated. The force constants among nearest $B-A$ and $B-B$ type atoms are same as nearest $A-B$ and $A-A$ atoms respectively.
\bea
\phi_{xx}(0,~1)&=& \phi_{xx}(1,~0)=-\gamma,\nonumber \\
\phi_{xy}(0,~1)&=& \phi_{xy}(1,~0)=\phi_{yx}(0,~1)= \phi_{yx}(1,~0)\nonumber\\
&=&\phi_{yy}(0,~1)=\phi_{yy}(1,~0)=0,\nonumber\\
\phi_{xx}(0,~2)&=& \phi_{xx}(0,~3)=\phi_{xx}(1,~4)=\phi_{xx}(1,~5)=-\frac{\gamma}{4},\nonumber\\
\phi_{xy}(0,~2)&=&\phi_{yx}(0,~2)=\phi_{xy}(1,~5)=\phi_{yx}(1,~5)=\frac{\sqrt {3} \gamma}{4},\nonumber\\
\phi_{xy}(0,~3)&=&\phi_{yx}(0,~3)=\phi_{xy}(1,~4)=\phi_{yx}(1,~4)=-\frac{\sqrt 3 \gamma}{4},\nonumber\\
\phi_{yy}(0,~2)&=& \phi_{yy}(0,~3)=\phi_{yy}(1,~4)=\phi_{yy}(1,~5)=-\frac{3 \gamma}{4},\nonumber\\
\phi_{xx}(0,~6)&=& \phi_{xx}(0,~9)=\phi_{xy}(0,~6)=\phi_{xy}(0,~9)=\phi_{yx}(0,~6)=\phi_{yx}(0,~9)=0, \nonumber\\
\phi_{yy}(0,~6)&=& \phi_{yy}(0,~9)=-\gamma_1,\nonumber
\eea
\bea
\phi_{xx}(0,~7)&=&\phi_{xx}(0,~8)=\phi_{xx}(0,~4)=\phi_{xx}(0,~5)=- \frac{3 \gamma_1}{4},\nonumber\\
\phi_{xy}(0,~7)&=& \phi_{yx}(0,~7)=\phi_{xy}(0,~5)=\phi_{yx}(0,~5)= \frac{\sqrt 3}{4}\gamma_1,\nonumber\\
\phi_{xy}(0,~8)&=& \phi_{yx}(0,~8)=\phi_{xy}(0,~4)=\phi_{yx}(0,~4)=-\frac{\sqrt 3}{4}\gamma_1 ,\nonumber\\
\phi_{yy}(0,~7)&=& \phi_{yy}(0,~8)=\phi_{yy}(0,~4)=\phi_{yy}(0,~5)=-\frac{\gamma_1}{4}.\nonumber
\eea
From sum rule in equation (\ref{eq:4}) the self force constants are
\bea
\phi_{xx}(0,~0) &=& -[\phi_{xx}(0,~1) + \phi_{xx}(0,~2) + \phi_{xx}(0,~3) + \phi_{xx}(0,~6) + \phi_{xx}(0,~9) +\nonumber\\
&& \phi_{xx}(0,~7) + \phi_{xx}(0,~8) + \phi_{xx}(0,~5) + \phi_{xx}(0,~4)] \nonumber \\
&= &\frac{3}{2} \gamma +3 \gamma_1. \nonumber
\eea
Similarly,
\be
\phi_{xy}(0,~0)=\phi_{yx}(0,~0)=0~~~\text{and}~~~
\phi_{yy}(0,~0)=\frac{3}{2} \gamma +3 \gamma_1.\nonumber
\ee
For the second atom the force constants can be calculated in the same way and the self force constants are
\be
\phi_{xx}(1,~1)=\phi_{yy}(1,~1)=\frac{3}{2} \gamma +3 \gamma_1~~\text{and}~~
\phi_{xy}(1,~1)=\phi_{yx}(1,~1)=0.\nonumber
\ee
Here (0~0) and (1~1) stand for A and B sublattices respectively. Now, the elements of the dynamical matrix are
\bea
D_{xx}(00) &= &\frac{1}{M}[\phi_{xx}(0,~0)+\phi_{xx}(0,~4) e^{-\imath(k_x a \frac{\sqrt 3}{2} + \frac{k_y a}{2})} + \phi_{xx}(0,~8) e^{\imath(k_x a \frac{\sqrt 3}{2} + \frac{k_y a}{2})} + \phi_{xx}(0,~5) e^{-\imath(k_x a \frac{\sqrt 3}{2} - \frac{k_y a}{2})} \nonumber \\
& &  + \phi_{xx}(0,~7) e^{\imath(k_x a \frac{\sqrt 3}{2}-\frac{k_y a}{2})} + \phi_{xx}(0,~6) e^{-\imath (k_y a)} + \phi_{xx}(0,~9) e^{\imath (k_y a)}] \nonumber \\
&= &\frac{1}{M}\left[\frac{3}{2}\gamma +3 \gamma_1 (1-\cos(k_x a \frac{\sqrt 3}{2}) \cos( \frac{k_y a}{2})\right].\nonumber\\
&=&D_{xx}(11)\nonumber
\eea
Let us call this element as $A_1$. Similarly,
\bea
D_{xy}(00)&=&D_{yx}(00)=D_{xy}(11)=D_{yx}(11)=-\gamma_1 \frac{\sqrt 3}{M} \sin(k_x a \frac{\sqrt 3}{2}) \sin( \frac{k_y a}{2})\equiv  B_1,\nonumber \\
D_{yy}(00)&=& D_{yy}(11)=\frac{1}{M}\left[\frac{3}{2} \gamma + \gamma_1 \{3-\cos(k_x a \frac{\sqrt 3}{2}) \cos( \frac{k_y a}{2}) - 2 \cos(k_y a)\}\right]  \equiv A_2, \nonumber\\
D_{xx}(01)&=&\frac{1}{M}\left[\phi_{xx}(0~~1) e^{-\imath(\frac{k_x a}{\sqrt 3})} +  \phi_{xx}(0~~2) e^{\imath(\frac{k_x a}{2 \sqrt 3} - \frac{k_y a}{2})} +
\phi_{xx}(0~~3) e^{\imath(\frac{k_x a}{2 \sqrt 3} + \frac{k_y a}{2})}\right] \nonumber \\
&=&-\frac{\gamma}{M}\left[e^{-\imath(\frac{k_x a}{\sqrt 3})} + \frac{1}{2}  e^{\imath(\frac{k_x a}{2 \sqrt 3})} \cos(\frac{k_y a}{2})\right]  \equiv C_1, \nonumber\\
&=&D^\ast_{xx}(10)\nonumber\\
D_{xy}(01)&=&D_{yx}(01)=D^\ast_{xy}(10)=D^\ast_{yx}(10)=-\imath \frac{\sqrt 3}{2M} \gamma e^{\imath(\frac{k_x a}{2 \sqrt 3})} \sin(\frac{k_y a}{2}) \equiv D_1,\nonumber\\
D_{yy}(01)&=&D^\ast_{yy}(10)=-\frac{3\gamma}{2M} e^{\imath(\frac{k_x a}{2 \sqrt 3})} \cos(\frac{k_y a}{2}) \equiv C_2. \nonumber
\eea
\vspace*{.3cm}
The secular determinant is\\ 
%\vspace*{.5cm}
$\left|\begin{array}{cccc}
\ A_1 - \omega^2 & B_1 & C_1 & D_1            \\
 B_1 & A_2 -\omega^2 & D_1 & C_2              \\
 C^\ast_1 & D^\ast_1 & A_1 - \omega^2 & B_1    \\
 D^\ast_1 & C^\ast_2 & B_1 & A_2 - \omega^2
\end{array}\right|. $\\ \\
%\vspace*{.5cm}
The hexagonal brillouin zone of graphene is shown in the fig.(16).
%\vspace*{.5cm}
\figgrphbz
In the first brillouin zone the symmetry points are $\Gamma(0,0)$, $M(2\pi/a\sqrt{3},0)$ and $K(2\pi/a\sqrt{3}, 2\pi/3 a)$. Let us now look for the eigen frequencies of lattice vibration at the above mentioned symmetry points of the brillouin zone within nearest neighbor coupling only, i.e. putting $\gamma_1=0$. Considering atomic mass to be unity, the eigenfrequencies at the $\Gamma$ point are
$\omega_1=0,~~\omega_2=\sqrt{3 \gamma},~~\omega_3=0,~~\omega_4=\sqrt{3 \gamma}$, at the $M$ point are $\omega_1=\sqrt{\gamma},~~\omega_2=\sqrt{2 \gamma},~~\omega_3=0,~~\omega_4=\sqrt{3 \gamma}$ and at the $K$ point are $\omega_1=0,~~\omega_2=\sqrt{3 \gamma}/2,~~\omega_3=\sqrt{3 \gamma}/2,~~\omega_4=\sqrt{3 \gamma}$. At $\Gamma$ point two of the eigenfrequencies are zero (acoustic branches). Other two non-zero values which correspond to optic branchesare are degenerate (which should be the case for a non-polar material). At $M$ point all four branches are nondegenerate and at $K$  point two of the branches are degenerate. The dispersion relations along symmetry line $\Gamma - M$ are 
\bea
\omega^2_1&=& \frac{3\gamma}{2}-\gamma \left[\frac{5}{4} + \cos(k_x a \frac{\sqrt 3}{2}\right], \\
\omega^2_2&=& \frac{3 \gamma}{2} + \gamma \left[\frac{5}{4} + \cos(k_x a \frac{\sqrt 3}{2}\right],\\
\omega^2_3&=& 0 ~~~\text{and}~~~\\
\omega^2_4&=& 3 \gamma.
\eea
The dispersion curves along all the symmetry directions are shown graphically in fig.(17).
\vspace{.4cm}
\figonepar
%\vspace*{2cm}
Though in case of diatomic square lattice one of the acoustic branches vanishes only along $\Gamma - X$ direction, in case of graphene one of the acoustic branches has totally zero frequencies over the entire symmetry directions when the next nearest neighbor coupling among the atoms is neglected. This indicates that to have quite satisfactory results over the entire brillouin zone, the n.n.n. coupling should be taken into account. In presence of both type of coupling, the eigen frequencies are $\omega_1=0,~~\omega_2=\sqrt{3 \gamma},~~\omega_3=0,~~\omega_4=\sqrt{3 \gamma}$ at the $G$ point and $\omega_1=\sqrt{\gamma + 6 \gamma_1},~~\omega_2=\sqrt{2\gamma + 6 \gamma_1},~~\omega_3=\sqrt{\gamma_1},~~\omega_4=\sqrt{3\gamma + 2\gamma_1}$ at the $M$ point.
The dispersion relations along all the symmetry lines in this case are shown in fig.(18).
\vspace*{.3cm}
\figtwopar
\section{Results and Discussions}
Comparing the vibration spectrum of triangular lattice with that of graphene, we see that one optical branch and one acoustic branch are degenerate at the $K$ point in graphene and both the acoustic branches are degenerate at $K$ point in triangular lattice. We also see that the full phonon spectrum over the entire brillouin zone for triangular lattice is obtained with nearest neighbor coupling among the atoms but for graphene one of the acoustic branches is totally suppressed over the whole brillouin zone in absence of second neighbor interaction. In the work by Falkovsky\cite{FALK} and earlier work by Woods and Mahan\cite{MAHAN} a detailed comparison with experimental result with graphene has been made. They describe the phonon dispersion with nearest and next nearest coupling among atoms.
Woods and Mahan\cite{MAHAN} have considered a two parameter (viz. $\alpha$ and $\beta$) model, where $\alpha$ is characteristic to two body central force between two neighboring atoms and it entirely depends on the separation between the atoms and $\beta$ is characteristic to three body force related to bending of two adjacent bonds. Thus, $\beta$ essentially gives the coupling between next-nearest neighbor atoms via the neighboring atom. Hence though carbon atoms in graphene are bonded covalently, we have used an effective coupling between the second neighboring atoms (considering coupling constant very small compared to first neighbor coupling constant) instead of exactly following the microscopic details like them. It follows that at high symmetry points viz. $\Gamma$, $M$, $K$ the analytical results are same in presence of nearest neighboring coupling only. In presence of two coupling constants we have got the analytic results at $\Gamma$ and $M$ points but not at $K$ point. But the numerical plot shows that two optical branches are degenerate at $\Gamma$ point, at $M$ point all four branches are non-degenerate and at $K$ point one acoustic and one optic branches are degenerate. These general features also match with the features obtained by Woods and Mahan\cite{MAHAN}. In Falkovsky's work\cite{FALK}, unlike equation (\ref{eq:5}), the atomic force constants have been calculated by imposing constraints arising from the point group symmetry of graphene and the general features of vibration spectrum at the high symmetry points are similar to results obtained here.
Thus, we see that one acoustic branch of the vibration spectra of monatomic and diatomic square lattices is suppressed along one direction of the brillouin zone whereas in graphene it is suppressed over the entire brillouin zone in the absence of second neighbor coupling. So, to have a full dispersion curve over the whole brillouin zone in graphene as well as in square lattices, second neighbor coupling should be taken into account.
\begin{center}
ACKNOWLEDGMENT
\end{center}
I am happy to thank Prof. S. G. Mishra for encouraging me and for his active help for this work, Dr. B. R. Sekhar and Mr. Nabyendu Das for fruitful discussions, Dr. Monodeep Chakraborty and Mr. Trilochan Bagarti for some technical helps.

\end{document}